# Assessing the Acceptance of Clinical Decision Support Tools using an Integrated Technology Acceptance Model

Soliman Aljarboa
*VU Business School, Victoria University,* Melbourne, Australia.
College of Business and Economics, Qassim University, Saudi Arabia
solimansalehm.aljarboa@students.vu.edu.au,
ssjrboa@qu.edu.sa

Shah Jahan Miah
*Newcastle Business School,
The University of Newcastle,*
Newcastle City Campus,
New South Wales
Australia
shah.miah@newcastle.edu

*Abstract*—With the medical development in recent decades, the multiplicity of different diseases and an increased number of patients, the use of an advanced healthcare information technology (HIT) such as a clinical decision support system (CDSS) has been of necessary to help general practitioners (GPs). CDSS may fail due to the failure to understand the factors influencing the GP's acceptance of CDSS. Identifying factors that promote acceptance of CDSS can be a vital aspect for its successful implementation. This study seeks to identify factors that influence the acceptance of CDSS in Saudi Arabia by GPs. This study relies mainly on the unified theory of acceptance and use of technology (UTAUT) which has been integrated with a task-technology fit (TTF) model and has applied a qualitative method to collect the data through using semi-structured interviews with 12 GPs. The study's results indicated that performance expectancy, effort expectancy, facilitating conditions, technology fit for the task, technology characteristics and task characteristics have all influenced the acceptance of CDSS. The results also indicated the need to extend the UTAUT model to investigate and explore other factors in GPs' acceptance of CDSS.

*Keywords*— CDSS, acceptance, UTAUT, TFF, GPs, Saudi Arabia

## I. INTRODUCTION

CDSS is one types of HIT that contribute significantly to diagnoses, dispensing of appropriate medications, providing recommendations and alerts, and providing relevant information that contributes to decision-making for the end-user [1]. CDSS is an emerging technology that needs more research and knowledge to reconcile and increase the interaction between the physician and CDSS in order to assist and support the former in adopting and using the system successfully [2]. CDSS implementation and adoption are hugely beneficial for medical professionals as well as for patients in obtaining filtered information. Whether from patient data or evidence-based information, that contributes to decision-making [3].

Using advanced systems such as CDSS contributes to the provision of more accurate medical decisions. In general, GPs have less up-to-date medical information in many disciplines compared with specialist physicians and consultants, so, GPs will be having a greater need in their medical practices for such systems, especially when the condition or complaint needs a specialist. In most cases, GPs are the first medical contact to meet and examine a patient [4]. so the results of the examination should be correct and accurate, and this can be helped by using an advanced system that contributes to the decision-making process, such as CDSS [5].

The investigation and discovery of factors that affect health systems acceptance are crucial and important for improving health care services and patient well-being. In addition, determining the factors affecting HIT will contribute significantly to developing fit and appropriate systems for the end-user [6].

## II. PROBLEM OF THE RESEARCH

Previous studies indicated the lack of studies about CDSS in Saudi Arabia context [7, 8] as well as in developing countries [9]. Additionally, a search of several databases, including IEEE Xplore, EBSCO and ScienceDirect, confirmed that there have been insufficient studies in area of CDSS. Thus, it is imperative to emphasis the need for in-depth research in order to discover factors that may affect the acceptance of CDSSs in Saudi Arabia. Previous research has demonstrated that the successful performance of new technology is profoundly connected to the level of understanding of the elements affecting users' expectations and intention to utilise it [10].

Acceptance of the CDSS is crucial in order to provide better health care services, since if the user does not accept the technology, the non-acceptance may affect negatively the health care and well-being of patients [1].

This research is based on UTAUT model as a main conceptual framework. Therefore, the research aims to extend and develop the model to explore and identify the factors that influence the acceptance of CDSS by GPs. It is considered simplistic as it focuses on limiting factors [11]. Shachak, et al. [11] indicated that it is vital for researchers to expand and develop a model to include different dimensions that contribute to understanding issues related to the implementation and use of HIT. This study has integrated

TTF model with the UTAUT model as the primary model of the research.

Improving the quality and safety of preventive and health care services continues to be a challenge in hospitals and primary health care centres [12]. This requires further research and studies in the development and improvement of HIT and knowledge of emerging issues and the needs on a scientific basis. There have been limited studies on the current situation, acceptance and use of HIT, so it is necessary to investigate both the positive and negative aspects of using CDSS in Saudi Arabia to generate knowledge and recommendations [8, 13].

## III. THE LITERATURE REVIEW

The use of CDSS has become commonplace in operating theatres where they guide physicians, for instance, to minimize treatment errors [14]. The CDSS provides several different tools such as automated warnings or alerts, reminders, graphic summaries of complicated details, models for reporting, relevant reference information, clinical guideline, and diagnostic test results [3].

Several studies have confirmed that the adoption of CDSS can help to present a significant role in supporting and guiding the decision-making [15]. However, different studies in the HIS showed that there are still various failures in CDSS implementation in some nations, that demand urgent consideration and investigation [16]. In particular, the uptake of CDSSs by GPs is limited [17]. In hospitals, it is observed that a diverse range of information system is working that belong to three types of clusters, namely administrative, clinical, and strategy. The electronic patient record and CDSS types systems belong to the clinical category and the clinical information system (CIS) [18], which is considered under the umbrella of clinical information system (CIS).

Kim, et al. [19] asserted in their study that the evaluation of CDSS is needed in other hospitals to improve drug safety and user satisfaction. Additionally, Williams, et al. [20] suggested combining more constructs with the UTAUT model from different theories in order to discover and identify other factors that influence technology acceptance. CDSS need to be developed and implemented using a technology that will reduce cost and increase quality [6]. Moreover, emphasis should be placed among developing nations to alleviate the suffering those who are less fortunate.

Devaraj, et al. [21] classified and recognised possible obstacles and facilitators which enhance and develop treatments and clinical practices by adopting CDSS. A list of obstacles and facilitators were classified under the four determinants of the UTAUT model. The authors indicated the following: "lack of time or time constraints, economic constraints (e.g., finance and resources), lack of knowledge of system or content, reluctance to use the system in front of patients, obscure workflow issues, less authenticity or reliability of the information, lack of agreement with the system, and physician or user attitude toward the system" [21].

## IV. SAUDI VISION 2030

Vision 2030 is a strategic plan by which the Saudi government aims to reduce the Kingdom's excessive dependence on oil and to diversify its sources of income through its available resources and capabilities [22].

The government of Saudi Arabia through *the Vision 2030* seeks to improve the economic, developmental, health, technical and educational situation in the country as well as to encourage the maximum exploration of other competitive resources that encourage economic growth. By identifying the needs of all economic sectors, the Vision ensures that economic developments in the country of Saudi Arabia are considered [22].

This initiative is aimed at improving the efficiency of healthcare services in all three levels (primary, secondary, and tertiary) by increasing the capacity of healthcare facilities. The program has an eye for international participation by relevant organizations that may consider investing in the country's healthcare system. The government intends to facilitate an increment of technology use in the health sector. This will be facilitated by investing in digital transformation and Information Technology infrastructure [22].

Adopting advanced health systems such as the CDSS are highly relevant to *Vision 2030* as it contributes to reducing medical errors and providing evidence-based information and knowledge, and also contributes significantly to providing high-quality health services.

## V. TECHNOLOGY ACCEPTANCE

Realising and understanding user perspective to use the technology has proved to be quite a critical issue of IT management and implementation [23]. With the introduction of new technology, firms usually aim to achieve a strategic advantage, decrease costs, and enhance performance. However, unless the technology is accepted by users, these aims do not seem to get real [23]. Various conceptual models were designed to explain the acceptance of users of technology, which has been operationalised as attitudes and positions towards it (i.e. technology) [24] as well as the behavioural intention to use the technology [25]. In order to improve health care services, physicians and other medical professionals should accept the HIT [26].

## VI. STUDY FRAMEWORK

### A. Unified Theory of Acceptance and Use of Technology (UTAUT)

The principal evaluation framework for the study is based on UTAUT model [25]. Figure 1 illustrates the model used in this study.

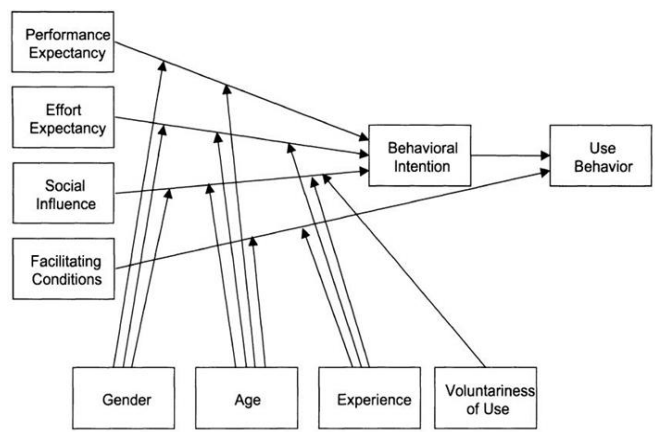

Fig. 1. UTAUT Model (Venkatesh et al. 2003)

The UTAUT model is based on the following eight different models:

The theory of reasoned action; the technology acceptance model; the motivational model; the theory of planned behaviour; a model combining the technology acceptance model and the theory of planned behaviour; the model of PC utilization; the innovation diffusion theory, and the social cognitive theory. The UTAUT model includes four different moderator variables which are: gender, age, experience, and voluntariness of use. It also includes four factors which are: performance expectancy, effort expectancy, social influence and facilitating conditions [25]. The determinants of UTAUT was defined as follows.

Performance Expectancy: The degree to which the user believes that the system will help to obtain necessary gains out for performing a job.

Effort Expectancy: The degree to which the system be easy to use for the user.

Social Influence: the degree to which the user feels that relevant others think they should use the system. This has been predicted by the model that, within the system of information, the user behaviour is affected by behavioural intention.

According to Venkatesh, et al. [25] these three determinants (social influence, performance expectancy and effort expectancy) have an immediate effect on users' behavioural intentions while facilitating conditions have a direct impact on user behaviour.

UTAUT model has been affirmed to be valuable and effective in a different field that focuses on evaluating the potential success of a new system and assists in understanding the factors of acceptance [27]. The flexibility and inclusiveness of the UTAUT one of the most important reason for selecting an effective model that integrated the most relevant factors from different eight models.

*B. Task-Technology Fit*

TTF model has been investigated in both information technologies and HIT [28, 29]. TTF examines the technology characteristics and the task characteristics to fit the requirements of the task to enhance the performance of the user [30].TTF states that user does not rely only on their feelings and perspectives to use the technology; however, it could identify the constructs of technology acceptance through a task- technology fit.

TTF model has two constructs: task characteristics and technology characteristics, which impact utilization and task performance [30]. TTF confirms that if technology provides characteristics that fit the requirements, then satisfactory performance will be achieved [31].

*C. The main framework of the study*

Several studies have combined TTF with the UTAUT model to investigate the technology acceptance factors [32, 33].

Furthermore, Khairat, et al. [1] provided a critical review of CDSS studies. They indicated that combining the models and frameworks would develop and enhance user acceptance to promote and assist the successful and useful adoption of CDSS. The authors stated that if the user did not accept the system, there would be a lack of use of the technological system and may threaten the healthcare and well-being of patients.

The combined between both UTAUT and TTF model provided several significant contributions to technology acceptance modelling. First, UTAUT has been produced in consequence of studying eight different acceptance theories then the most influencing factors were selected as the main components of UTAUT. According to Zhou, et al. [34], UTAUT was appropriate to identify user's attitudes to accept the system; however, the model did not adequately explain its task technology fit. Zhou, et al. [34] studied the adoption of mobile banking in China and noticed that the integrated model of TTF-UTAUT led to having better outcomes rather than the applying of a single framework or theory.

For better understanding, the problems related to user acceptance in healthcare, TTF need to be incorporated by evaluation frameworks with the factors of user acceptance. TTF considered being a significant addition for understanding the needs of user acceptance.

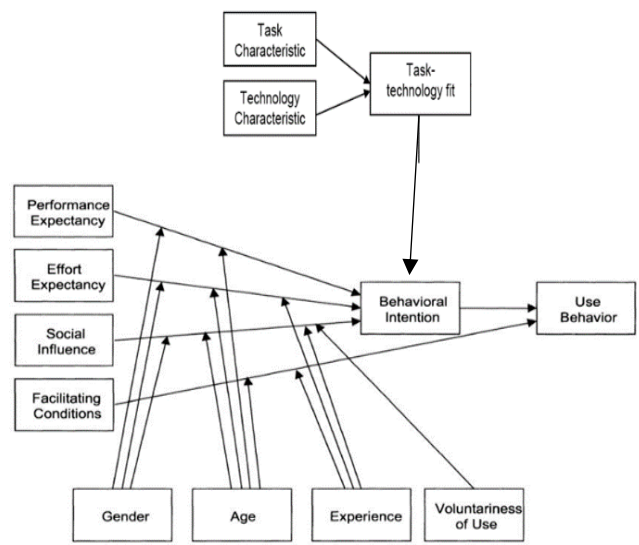

Fig. 2. Model of Integration of UTAUT and TTF

VII. STUDY METHODOLOGY

This study applied a qualitative method to collect data by conducting semi-structured interviews. We use a semi-structured interview approach in order to obtain more data by asking questions and inquiries based on open-ended questions. The questions were prepared in advance based on the constructs of UTAUT and TTF model. The discussion with participants has extended via the control of the interview by the researcher, where the inquiries have been continued by more questions to have more understanding and comprehensive answers.

The semi-structured interview helped us to obtain detailed information on the factors influencing acceptance of a CDSS for healthcare in Saudi Arabia. In this study, twelve GPs who are currently working in various healthcare establishments or working at the local hospitals in Saudi Arabia has participated. Qualitative research focuses on "exploring and understanding the meaning individuals or groups ascribe to a social or human problem" [35, P. 4 ]. A qualitative approach provides rich and abundant data that contributes a greater knowledge of the issue or a problem of the study [36]. Thus, using qualitative research allows for a deep exploration and understanding of participants' individual perspectives. With the use of semi-structured interviews, participants freely provide their viewpoints. This approach sufficiently satisfies the objectives of the research.

VIII. DATA ANALYSIS

Thematic analysis was used to analyse the data gathered from the interviews to know more about their experiences, perspectives, and attitudes [37]. Such analyses of data contribute to the formation of theories through a set of procedures that help to generate facts and factors [38].

The data was mainly analysed using the NVivo software, following the six-step phases of employed thematic analysis established by Braun and Clarke [37].

All transcripts were collected in folders dedicated in the NVivo program for the purpose of objective analysis for a more in-depth review and study of the content, understanding the perceptions of the participants in the study and exploring the factors that affect physicians' acceptance of the system. Once the NVivo data has been manually formatted data in various aspects [39], it is easy to use to organise and classify data in an orderly and coordinated manner. It also makes it easier for the researcher to delve deeper into the data [40]. Further, NVivo allows a researcher to specify sentences, phrases, and words in the content. NVivo also contains features that help to easily identify and arrange the main and sub-themes [41].

IX. DISCUSSION AND RESULTS

Twelve GPs from different hospitals and primary health care centres in Saudi Arabia participated in this study. Each interview lasted approximately 30-50 minutes. The participants contributed significantly to providing valuable information about the factors that affect the acceptance of CDSS.

**Findings in relation to the framework**

*A. Performance expectancy*

The factor of Performance Expectancy has been used in different models of acceptance technology, including UTAUT [25]. In this study, the Performance Expectancy factor is a significant determinant of the effect on GPs' acceptance of intention to use CDSS: all interviewees acknowledged the importance of this factor in their acceptance and accreditation in the use of CDSS. Time, alerts, accuracy, and reduced errors were identified as the main circumstances underlying Performance Expectancy contributing to the admission process. It was identified as an important and fundamental determinant of acceptance of technologies in many fields of science such as health [42].

Most GPs indicated the extent of the CDSS's contribution to saving time and obtaining information at the required time .

SA6 said, *"CDSS contribute to obtaining the information in real-time, meaning that if you want to request a patient file from the hospital or even from another hospital, it would be quick to get his or her information".*

*B. Effort expectancy*

The study confirms that Effort Expectancy is one of the main factors that contribute to the acceptance by GPs to their use of the CDSS .

There is clear evidence from several studies in different fields of information technology (IT) that promote the affirmation that Effort Expectancy has a positive impact on the behavioural intention to use HIT [43]. In this research, the Effort Expectancy is a significant factor of the effect on GPs' acceptance of intention to use CDSS. In the interviews with GPs, most indicated that the CDSS would be easy to use due to their good technology background. One of the participants,

SA7 said, *"I do not think there will be any difficulty in using the system; however, with the frequent use of the system, it will be easier ".*

*C. Social influence*

Social influence is defined as the degree to which a user accepts or thinks that significant others would affect him or her to use the system. Most interviewees indicated that people who might have an influence on their work had no impact on their decision to use CDSS. However, several GPs indicated that co-workers could play an influential role in their use of CDSS. There are numerous studies in several fields and the area of HIT; the research has indicated that social influence does not affect the intention to use IT and applications [17, 44].

*D. Facilitating conditions*

The study confirmed the importance and role of Facilitating Conditions as a key factor in influencing GPs' acceptance and use of the CDSS. In the interviews, the participants agreed on the importance of the following elements: training [11]; technical support, and updating the IT system. Several studies have mentioned the importance and the positive role of Facilitating Conditions in influencing the use of IT [45]. and health information systems in particular [43].

SA1 mentioned that *"Training is one of the most important things in discovering and getting to know the system, with training CDSS will be easier to use"*

*E. Technology Characteristics*

Interviews with GPs demonstrated the role of the characteristics of technology as an influencing factor in their acceptance of intention to use CDSS. This factor has been added from the TTF model to the main model approved in the study, as it has been considered in many studies regarding issues of acceptance of IT [45].

IT is considered as both tools and systems employed by users to complete their tasks. The participants indicated the importance of the role of technology characteristics that it includes fast internet and modern computers which help and contribute effectively and influential to the TTF and thus contribute to the acceptance of the CDSS. Most GPs commented that the presence and availability of such features in CDSS contributes and helps them to save time and to perform the task in a timely manner. This also helps the patient by reducing waiting time as well as the time for diagnosis and reduction of consultation times.

*F. Task characteristics*

The Task Characteristics factor is another factor that has been added to the primary model of the study after merging the TTF and UTAUT Models. Most participants indicated the extent of the Task Characteristics' role and its positive impact on the acceptance of intention to use CDSS. Most participants considered that the daily work has degrees of difficulty as well as complexity, as in some cases the patient's diagnosis requires more experience and practice of the GP, who may conduct further tests in some cases. The Task Characteristics factor has been examined in several studies in relation to the acceptance of IT [46].

*G. Task-Technology Fit (TTF)*

TTF determinant means a belief that the CDSS matches the GPs' task needs and their capabilities [47]. In other words, TTF determines which technologies benefit tasks while the features present the tools that sufficiently meet user requirements. The GPs confirmed that the nature of the medical work and their daily medical practices are closely aligned to and fit with CDSS, which will help them to provide better medical services and care. TTF has been used in many different studies in the field of HIT[48]. Most GPs stated that CDSS would achieve their desired goals and that TTF would have a positive impact on the acceptance of the CDSS.

## I. FUTURE RESEARCH AND CONCLUSION

This study could be developed further by investigating the factors that influence GPs' and other medical workers' acceptance of CDSS. Most GPs indicated the influences of UTAUT and TTF model factors, except the social influence factor. They also indicated that there are other factors that also influence the acceptance of CDSS. This leads to the conclusion that a more in-depth study is required which would discover the factors affecting the acceptance of the CDSS through reliance on qualitative research. This study proposes to conduct more interviews in two stages. The first stage is based on convergent interviewing with about 12 GPs intended to discover and determine the factors that affect user acceptance. The second phase, consisting of more in-depth interviews, follows the case study approach and investigates the factors that were discovered in the first phase and investigate in greater detail the factors of the UTAUT and TTF models (interviews with 30-40 GPs). This approach will contribute towards a complete understanding and richly detailed data about the factors that influence the acceptance of the system by the GPS and contribute to confirming the extent of the research results. The authors intend to present a more in-depth study using the qualitative approach to discover new factors affecting the GPs' acceptance of CDSS.

This research aimed to investigate the factors that influence CDSS acceptance among GPs in healthcare in Saudi Arabia. We found that the integration of the UTAUT and TTF models provided a better measure for defining the usage behaviour. The results showed that all factors of both UTAUT and TTF influence the acceptance of CDSS by GPs. The aim of this research is to build a real healthcare decision system [49]. This research provides an opportunity for future research to develop a new framework that influence the acceptability of any new HIS design (for instance, using design science research [50, 51, 52]).